\documentclass[reprint,
superscriptaddress,
preprintnumbers,
 amsmath,amssymb,
 aps,
prd,
longbibliography,
]{revtex4-2}
\usepackage{CJKutf8}
\usepackage{multirow}
\usepackage{natbib}
\usepackage{mathtools}
\usepackage{xcolor}
\usepackage{graphicx}
\usepackage{dcolumn}
\usepackage{bbm}
\usepackage{hyperref}

\usepackage{tikz-feynman}
\usepackage{slashed}

\DeclareMathAlphabet\mathbfcal{OMS}{cmsy}{b}{n}

\newcommand{\ra}{\rangle}

\newcommand{\la}{\langle}

\newcommand{\eq}{\begin{equation}}
\newcommand{\eqe}{\end{equation}}
\newcommand{\eqa}{\begin{eqnarray}}
\newcommand{\eqae}{\end{eqnarray}}
\newcommand{\nn}{\nonumber}
\newcommand{\bn}{\begin{enumerate}}
\newcommand{\en}{\end{enumerate}}
\def\beq#1\eeq{\begin{align}#1\end{align}}

\newcommand{\eqc}[1]{Eq.(\ref{#1})}

\def\CO{{\mathcal O}}

\def\CM{{\mathcal M}}

\def\CO{{\mathcal O}}

\newcommand{\bfig}{\begin{figure}}
\newcommand{\efig}{\end{figure}}

\def\bl#1\el{\begin{align} #1 \end{align}}
\def\bg#1\eg{\begin{gather} #1 \end{gather}}

\def\bld#1\eld{\begin{aligned} #1 \end{aligned}}
\def\bgd#1\egd{\begin{gathered} #1 \end{gathered}}

\newcommand{\ket}[1]{|{#1}\rangle}

\newcommand{\sbra}[1]{ [{#1} |}
\newcommand{\sket}[1]{ | {#1} ]}

\def\topbotatom#1{\hbox{\hbox to 0pt{$#1\bot$\hss}$#1\top$}} \newcommand*{\topbot}{\mathrel{\mathchoice{\topbotatom\displaystyle} {\topbotatom\textstyle} {\topbotatom\scriptstyle} {\topbotatom\scriptscriptstyle}}}

\def\bubatom#1{\hbox{\hbox to 1pt{$#1($\hss}$#1)$}}  \newcommand*{\bub}{\mathrel{\mathchoice{\bubatom\displaystyle} {\bubatom\textstyle} {\bubatom\scriptstyle} {\bubatom\scriptscriptstyle}}}

\renewcommand{\tt}{\texttt}

\newcommand{\RN}[1]{%
  \textup{\uppercase\expandafter{\romannumeral#1}}%
}

\newcommand{\be}{\begin{equation}}
\newcommand{\ee}{\end{equation}}
\newcommand{\ba}{\begin{align}}
\newcommand{\ea}{\end{align}}
\newcommand{\bi}{\begin{itemize}}
\newcommand{\ei}{\end{itemize}}

\let\a=\alpha \let\b=\beta \let\g=\gamma  \let\e=\epsilon
\let\z=\zeta  \let\th=\theta  \let\k=\kappa
\let\l=\lambda \let\m=\mu \let\n=\nu \let\x=\xi   
     
\let\w=\omega      \let\L=\Lambda

\makeatletter
\newcommand*{\Rom}[1]{\expandafter\@slowromancap\romannumeral #1@}
\makeatother

\makeatletter
\newcommand*{\rom}[1]{\expandafter\romannumeral #1}
\makeatother

\let\nn=\nonumber

\def\beq#1\eeq{\begin{align}#1\end{align}}

\begin{document}
\begin{CJK*}{UTF8}{mj}
\preprint{QMUL-PH-22-20}
\title{Quantum corrections to frame-dragging in scattering amplitudes
}
\author{Jung-Wook Kim (김정욱)}
\email{jung-wook.kim@qmul.ac.uk}
\affiliation{Centre for Theoretical Physics, Department of Physics and Astronomy,\\Queen Mary University of London, Mile End Road, London E1 4NS, United Kingdom}
\affiliation{Kavli Institute for Theoretical Physics, University of California, Santa Barbara, California 93106, United States}

\begin{abstract}
Frame-dragging effect manifests itself as polarization direction rotation when linearly polarized electromagnetic/gravitational wave scatters from a spinning point source through gravitational interactions, an effect also known as the gravitational Faraday rotation. Treating general relativity as an effective field theory, the Faraday rotation angle and its quantum corrections are computed using scattering amplitudes. While the classical rotation angle is universal as expected from the equivalence principle, the quantum corrections are found to be different between electromagnetic and gravitational waves, supporting earlier studies that some formulations of the equivalence principle may need reformulation in the quantum regime.
\end{abstract}

\maketitle
\end{CJK*}

\section{Introduction}
Quantum gravity is incomplete in the sense that no known theory remains valid at all energy scales. However, it is known that any theory of gravitons (defined as self-interacting massless spin-2 particles) obeying certain physical principles such as unitarity, Lorentz invariance, and locality must reduce to general relativity (GR) at sufficiently low energy or curvature scales~\cite{Weinberg:1965rz,Feynman:1996kb,Benincasa:2007xk}. Moreover, GR as a quantum theory is a perfectly valid effective theory which gives consistent predictions below the cutoff scale, where the cutoff is conjectured to be of the order of the Planck mass~\cite{Donoghue:1993eb,Donoghue:1994dn,Burgess:2003jk}. Therefore, it is still possible to make definite predictions in quantum gravity which do not depend on the ultraviolet completion of the theory. The aim of this study is probing quantum corrections to frame-dragging, also known as the Lense-Thirring effect.

Frame-dragging is a feature of GR for which its Newtonian counterpart does not exist. Thus, the study of quantum corrections to frame-dragging is a study of quantum GR, which is qualitatively different from studies of quantum effects in Newtonian gravity; the former cannot be faithfully studied from the latter. This makes quantum frame-dragging an interesting subject of study as usual studies of quantum effects focus on phenomena having Newtonian analogs~\cite{Donoghue:1993eb,Donoghue:1994dn,Khriplovich:2002bt,Bjerrum-Bohr:2002gqz,Bjerrum-Bohr:2013bxa,Bjerrum-Bohr:2014zsa,Bjerrum-Bohr:2016hpa}.

Frame-dragging is sourced by angular momentum, and localized distribution of angular momentum can be approximated as a point source with spin. Spin effects in quantum GR have already been studied in the literature; as corrections to the metric~\cite{Donoghue:2001qc,Bjerrum-Bohr:2002fji,Holstein:2006ud,Holstein:2006ge}, or as corrections to the potential~\cite{Holstein:2008sx,Frob:2016xte}. However, both the metric and the potential are gauge-dependent quantities in GR, which renders them incomplete as observable effects. Moreover, the spins of the point sources considered in the works were quantum spins having values of $s = 1/2$ or $s = 1$. There is a conceptual gap between quantum spin (having values $s = 1/2, 1, 3/2, \cdots$) and classical spin
, and closure of the gap seems necessary to make connections with potentially realizable experiments; the frame-dragging effect sourced by quantum spins would be too small to have any observable consequences. Fortunately, methods to circumvent both limitations have already been studied in the literature.

The limitation of gauge-dependent quantities can be overcome by studying scattering observables measured at asymptotic infinity, which are free of gauge ambiguities. The approach has already been used to argue breakdown of the equivalence principle in the quantum realm~\cite{Bjerrum-Bohr:2014zsa,Bjerrum-Bohr:2016hpa,Bai:2016ivl,Chi:2019owc}. A scattering observable that is a direct consequence of frame-dragging effect is the gravitational Faraday rotation (GFR), where polarization direction of linearly polarized light (or gravitational wave) is rotated due to the ambient gravitomagnetic field sourced by angular momentum~\cite{Ishihara:1987dv, Nouri-Zonoz:1999jls, Sereno:2004jx, Brodutch:2011eh, Farooqui:2013rga, Shoom:2020zhr, Deriglazov:2021gwa, Chakraborty:2021bsb, Li:2022izh}. The rotation angle can also be computed using scattering amplitudes of quantum field theory~\cite{Chen:2022yxw}, which makes it an ideal observable for studying quantum frame-dragging.

The conceptual gap between quantum and classical spin has been studied in the context of scattering amplitude approaches to the classical gravitational two-body problem~\cite{Donoghue:1994dn, Bjerrum-Bohr:2002gqz, Bjerrum-Bohr:2013bxa, Neill:2013wsa, Cheung:2018wkq, Bern:2019nnu, Bern:2019crd, Bern:2021dqo}. While it is known that certain classical spin effects can be captured by quantum spins~\cite{Holstein:2008sx, Vaidya:2014kza, Maybee:2019jus, Chiodaroli:2021eug, FebresCordero:2022jts}, the description fails to capture terms that are affected by identities dependent on spin representations of the spinning particle, such as trace identities. Therefore, higher-spin ($s>2$) particle descriptions are required to capture the full dynamics of classical spinning particles~\cite{Guevara:2018wpp,Chung:2018kqs,Arkani-Hamed:2019ymq,Aoude:2020onz,Bern:2020buy}, which has met with great success in determining the classical two-body dynamics in GR~\cite{Guevara:2018wpp, Chung:2018kqs, Guevara:2019fsj, Chung:2019duq, Bern:2020buy, Kosmopoulos:2021zoq, Aoude:2021oqj, Chen:2021qkk, Aoude:2022trd, Bern:2022kto, Aoude:2022thd, Menezes:2022tcs}. In this work the semiclassical Compton amplitude~\cite{Chen:2021qkk}, which is defined as the Compton amplitude in the classical spin limit ($s \to \infty$ with $S = s\hbar$ fixed), is used. Note that amplitude computations at tree level~\cite{Guevara:2018wpp,Chung:2018kqs,Guevara:2019fsj,Arkani-Hamed:2019ymq,Chung:2019duq,Cangemi:2022abk} and one-loop level~\cite{Chung:2019yfs,Aoude:2022trd,Aoude:2022thd,Menezes:2022tcs} with spins in the classical spin limit have been checked against corresponding classical computations, and found to reproduce classical spin dependence to all orders in spin, which can be viewed as evidence for the validity of the approach.

The GFR angle $\a$ and its leading quantum corrections at linear order in spin are computed to be
\begin{subequations} \label{eq:Faraday_rot_ang_0}
\bl
\a_X (b) &= \frac{5 \pi G^2 m^2 (\hat{k} \cdot \vec{a})}{4 b^3} + c_{X} \frac{G^2 m \hbar (\hat{k} \cdot \vec{a})}{\pi b^4} \,, \label{eq:Faraday_rot_ang_1}
\el
where $m$ is the mass of the gravitating point source, $\vec{a} = \vec{S} / m$ is its spin-length vector, $b$ is the impact parameter, $\hat{k} = \vec{k} / \w$ is the propagation direction of the electromagnetic/gravitational wave, and $X = \g, h$ denotes particle species. 
\bl
\begin{aligned}
c_\g &= - \frac{994}{15} + 32 \log \frac{b}{b_0} & && & \text{photon}
\\ c_h &= - 60 + 24 \log \frac{b}{b_0} & && & \text{graviton}
\end{aligned} \label{eq:Faraday_rot_ang_2}
\el
\end{subequations}
The leading classical term of \eqc{eq:Faraday_rot_ang_1} agrees with known geometric optics limit results~\cite{Ishihara:1987dv, Li:2022izh, Chen:2022yxw}. Similar to the deflection angle~\cite{Bjerrum-Bohr:2014zsa,Bjerrum-Bohr:2016hpa,Bai:2016ivl,Chi:2019owc}, the quantum corrections to the GFR angle differ for different particle species.

In terms of the deflection angle $\th = \frac{4Gm}{b}(1 + \CO(\frac{Gm}{b}))$, the GFR angle can be written as
\bl
\bld
\a_X (\th) &= \frac{5 \pi}{256} (\hat{k} \cdot \vec{\chi}) \, \th^3 + \tilde{c}_X \frac{m_{\text{Pl}}^2}{m^2} (\hat{k} \cdot \vec{\chi}) \, \th^4 \,,
\\ \tilde{c}_X &= \left\{
\begin{aligned}
\tilde{c}_\g &= - \frac{497}{240} - \log \frac{\th}{\th_0} & && & \text{photon}
\\ \tilde{c}_h &= - \frac{15}{8} - \frac{3}{4} \log \frac{\th}{\th_0} & && & \text{graviton}
\end{aligned}
\right. \label{eq:Faraday_rot_ang_3}
\eld
\el
where $\vec{\chi} = \vec{S}/Gm^2$ is the dimensionless spin parameter and $m_{\text{Pl}} = \sqrt{\hbar / 8\pi G}$ is the reduced Planck mass. This expression has the advantage that all quantities on both sides can (in principle) be directly measured at asymptotic infinity, and that ill-definedness of $b$ at the wave-packet scale can be avoided~\footnote{A particle's position is ill-defined at the wave-packet scale. Another approach would be to reverse-engineer $b$ from the deflection angle, which is the approach of \eqc{eq:Faraday_rot_ang_3}.}. It would be interesting to study whether measurements at asymptotic infinity open up loopholes that can be used to bypass the known no-go results for experimental observation of quantum gravity effects~\cite{Rothman:2006fp,Dyson:2013hbl}.


\section{Computation setup}
Consider a massive-massless $2 \to 2$ scattering $p_1 k_2 \to p_1' k_2'$, where $p$ denotes massive momenta with mass $m$ and $k$ denotes massless momenta, while $q = k_2' - k_2 = p_1 - p_1'$ is the transfer momentum. The massive particle has spin $S^\m$, causing GFR on massless particle's polarization through frame-dragging.

The GFR angle $\a$ is obtained from the difference of phase gain between two helicity states of the massless particle. The difference in phase gain is identified with the difference in eikonal phases $\delta$ of helicity-preserving scattering amplitudes~\cite{Chen:2022yxw},
\bl
\a := - \frac{\delta_{+} - \delta_{-}}{2|h|} \,, \label{eq:alpha_def}
\el
where the eikonal phase $\delta_{\pm}$ is the logarithm of the scattering amplitude in impact parameter space (IPS)~\cite{Cheng:1969eh, Abarbanel:1969ek, Levy:1969cr, Amati:1987wq, Amati:1987uf, Kabat:1992tb, Akhoury:2013yua, KoemansCollado:2019ggb}.
\bl
\bld
e^{i \delta_{\pm}} &:= \int \frac{d^D q}{(2\pi)^D} P(q) \langle p_1' , k_2' (\pm) | S | p_1 , k_2 (\pm) \rangle \,,
\\ P(q) &= e^{- i b \cdot q} \hat{\delta}[q \cdot (p_1 + p_1')] \hat{\delta}[q \cdot (k_2 + k_2')] \,.
\eld \label{eq:eik_phase_def}
\el
Here $\hat{\delta}(x) = 2 \pi \delta (x)$, and $\pm$ is the massless particle's helicity. The classical terms of the eikonal phase satisfy exponentiation under the adopted projection condition $P(q)$~\cite{Brandhuber:2021eyq,Cristofoli:2021jas}. Only the nonanalytic terms of the scattering amplitude contribute to finite $b$ eikonal phase, which are constructible from unitarity-based methods~\cite{Bern:1994zx, Bern:1994cg, Bern:1997sc}.

\begin{figure}
\includegraphics{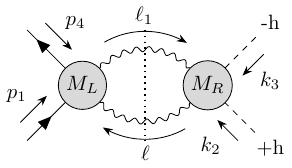}
\caption{The two-graviton cut for the one-loop amplitude $M_4 (p_1,k_2^{+h},k_3^{-h},p_4)$. Solid lines denote massive semiclassical spinning particle, dashed lines denote massless helicity $h$ particle, and wavy lines denote internal exchanged gravitons. $M_L$ is the semiclassical gravitational spinning Compton amplitude [\eqc{eq:Compton}], and $M_R$ is the two-graviton--two-helicity-$h$ amplitude [\eqc{eq:4grav}]. Figure drawn with package \textsc{TikZ-Feynman}~\cite{Ellis:2016jkw}.}
\label{fig:cut}
\end{figure}

The relevant $S$-matrix elements $iM$ can be obtained from the $4 \to 0$ amplitude $M_4 (p_1,k_2^{+h},k_3^{-h},p_4)$ using crossing symmetry. The relevant tree-level amplitudes were found to not contribute to GFR~\cite{Chen:2022yxw},
\bl
\bld
M_4^{(0)} &= M_{4, s=0}^{(0)} \times \exp \left( \frac{i \sbra{2} A \ket{3}}{\sbra{2} p_1 \ket{3}} \right) \,,
\\ \frac{M_{4, s=0}^{(0)}}{(\k/2)^2} &= \left\{ 
\begin{aligned}
& \phantom{a} - \frac{\sbra{2} p_1 \ket{3}^2}{t} && h = 1
\\ &\frac{\sbra{2} p_1 \ket{3}^4}{(s-m^2)(u-m^2)t}  && h = 2
\end{aligned}
\right.
\eld \label{eq:treeComp}
\el
where $s = (p_1 + k_2)^2$, $t = (k_2 + k_3)^2$, $u = (p_1 + k_3)^2$, and $\k = \sqrt{32 \pi G}$. The case $h=2$ corresponds to the semiclassical spinning Compton amplitude for Kerr black holes valid to quartic order in spin~\cite{Guevara:2018wpp,Guevara:2019fsj,Bautista:2019tdr,Aoude:2020onz,Chen:2021qkk}, which also enters the two-graviton cut in Fig.\ref{fig:cut} as $M_L$. The vector $A^\mu = (k_2 + k_3)_\nu S^{\m\n} (p_1)$ was introduced to simplify the notations~\footnote{The vector $A^\m$ is manifestly invariant under the shift $S^\m \to S^\m + \xi q^\m / q^2$ of Ref.\cite{Bern:2022kto} where $q^\m = - k_2^\m - k_3^\m$. The invariance of the amplitude under the shift is violated because the terms dependent on $p_1 \cdot S = 0$ are dropped when products of Levi-Civita tensors are reduced, while $p_1 \cdot q \neq 0$. The exact shift invariance would be restored if all $p_1 \cdot S$ were kept. Further discussions can be found in updated versions of Ref.\cite{Chen:2022yxw}.}. The spin tensor $S^{\m\n} (p_1)$ is the Lorentz generator $J^{\m\n}$ restricted to the little group space of the incoming particle of momentum $p_1$, satisfying the covariant spin supplementary condition $p_{1\m} S^{\m\n} = 0$~\footnote{While the canonical algebra satisfied by $S^{\m\n}$ inherited from $J^{\m\n}$ is tied to a \emph{different} spin supplementary condition~\cite{Steinhoff:2011sya}, the covariant one is chosen because the little group space is defined in the particle's rest frame. Identifying spin operators as the rotation generators in the rest frame of the particle is equivalent to choosing the covariant spin supplementary condition.} and related to the spin vector by $S^{\m\n} = - \frac{1}{m} \e^{\m\n\a\b} p_{1\a} S_\b$. The spin vector is the mass rescaled Pauli-Lubanski pseudovector $S^\m = \frac{-1}{2m}\e^{\m\a\b\g}p_{1\a} J_{\b\g}$~\cite{Bern:2020buy,Kosmopoulos:2021zoq,Chen:2021qkk}. Only the terms up to linear order in spin are considered, but the extension to higher orders would be straightforward.

The relevant terms of the one-loop amplitude $M_4^{(1)}$ can be constructed from the two-graviton cut in Fig.\ref{fig:cut}. The cut has nonvanishing contributions only when internal graviton states have opposite helicities~\cite{Bjerrum-Bohr:2016hpa}, e.g. $\ell^+$ and $\ell_1^+$. For this helicity configuration the subamplitudes are given as~\footnote{The spinor notations follow that of Ref.\cite{Chung:2018kqs} where incoming positive helicity states are associated with $\sket{p}$ spinors. Normalization for the Levi-Civita tensor is $\e^{0123} = +1$.}
\bl
M_L^{\ell^+,-\ell_1^-} &= \frac{\left( - {\k}/{2} \right)^2 \sbra{\ell} p_1 \ket{\ell_1}^4}{-4 t (k_2 \cdot \ell) (k_2 \cdot \ell_1)} \exp \left( \frac{i \sbra{\ell} A \ket{\ell_1}}{\sbra{\ell} p_1 \ket{\ell_1}} \right) \,, \label{eq:Compton}
\\ M_R^{-\ell^-,\ell_1^+} &= \frac{\left( - {\k}/{2} \right)^2 \sbra{\ell_1}2\ket{\ell}^4}{-4 t (k_2 \cdot \ell) (k_2 \cdot \ell_1)} \left( \frac{\la 3 \ell \ra}{\la 2 \ell \ra} \right)^{2h} \,, \label{eq:4grav}
\el
where $M_L = M_{\text{Comp}} (p_1, \ell^{+2}, -\ell_1^{-2}, p_4)$ is the semiclassical spinning Compton amplitude \eqc{eq:treeComp}, and  $M_R = M_{XhhX} (k_2^{+h}, \ell_1^{+2}, -\ell^{-2}, k_3^{-h})$ is the two-graviton--two-helicity $h$ amplitude valid for $h \le 2$. The cases $h = 1, 2$ are considered, which correspond to the Einstein-Maxwell theory ($h = 1$) and GR ($h = 2$).

Using unitarity-based methods~\cite{Bern:1994zx, Bern:1994cg, Bern:1997sc}, the one-loop integrand for $M_4 (p_1,k_2^{+h},k_3^{-h},p_4)$ is constructed from the two-graviton cut as~\footnote{The use of four dimensional helicity scheme~\cite{Bern:1991aq,Bern:2002zk} in dimensional regularisation of the loop integral is implicit, since four dimensional spinor variables were used to construct the integrand.}
\bl
i M_4^{(1)} &= \int \hskip-4pt \frac{d^D \ell}{(2\pi)^D} \frac{\left[ M_L^{\ell^+,-\ell_1^-} \hskip-2pt M_R^{-\ell^-,\ell_1^+} \right] + (\ell \leftrightarrow - \ell_1)}{2 \ell^2 \ell_1^2} \,, \label{eq:cut_integrand_1}
\el
where the ``numerator'' $M_L \times M_R$ captures all momentum dependence of the full one-loop integrand subject to the cut conditions $\ell^2 = 0$ and $\ell_1^2 = (\ell - k_2 - k_3)^2 = 0$, which determines the nonanalytic $t$ dependence of the full amplitude relevant to long-distance physics~\cite{Bjerrum-Bohr:2013bxa}. The second term $(\ell \leftrightarrow - \ell_1)$ in the numerator of \eqc{eq:cut_integrand_1} denotes opposite helicity configuration for the gravitons. 
Explicit evaluation of the first term $M_L^{\ell^+,-\ell_1^-} M_R^{-\ell^-,\ell_1^+}$ for $n$th order in spin yields
\bl
\bld
\left. \frac{M_L M_R}{(\k/4)^4}\right|_{\CO(S^n)} \hskip -9pt &= \frac{i^n}{n!} \frac{[\text{tr}_-(\slashed{k_2}\slashed{\ell_1}\slashed{p_1}\slashed{\ell})]^{4-n} [\text{tr}_-(\slashed{k_2}\slashed{\ell_1}\slashed{A}\slashed{\ell})]^{n}}{t^2 (p_1 \cdot \ell)(p_1 \cdot \ell_1)(k_2 \cdot \ell)(k_2 \cdot \ell_1)}
\\ &\phantom{=a} \times \frac{1}{\sbra{3} p_1 \ket{2}^{2h}} \left( \frac{\text{tr}_-(\slashed{k_2}\slashed{\ell}\slashed{k_3}\slashed{p_1})}{2(k_2 \cdot \ell)} \right)^{2h} \,,
\eld \label{eq:loop_integrand_1}
\el
where $\text{tr}_{-} (\slashed{a} \slashed{b} \cdots) = \text{tr} (\frac{1 - \gamma_5}{2} \slashed{a} \slashed{b} \cdots)$. The vector $A^\m$ is expanded on the basis formed from external momenta
\bl
A^\mu = a_1 p_1^\m + a_2 k_2^\m + a_3 k_3^\m + a_4 \e^{\a\b\g\m}p_{1\a} k_{2\b} k_{3\g} \,, \label{eq:A_exp}
\el
where the coefficients $a_i$ are rational functions of Mandelstam invariants and spin-vector Lorentz invariants
\bl
\{ \, (k_2 \cdot S) \,,\, (k_3 \cdot S) \,,\, \e_{\a\b\g\m} p_1^\a k_2^\b k_3^\g S^\m \, \} \,. \label{eq:spin_Lor_inv}
\el
All loop-momentum-dependent Levi-Civita contractions take the form $\e_{\a\b\g\m} p_1^\a k_2^\b k_3^\g \ell^\m$ under the expansion \eqc{eq:A_exp}. Any expression linear in this Levi-Civita contraction vanishes after loop-momentum integration~\cite{Chi:2019owc},
so they can be discarded before performing reduction to master integrals.

Reduction to master one-loop integrals was performed using the package \textsc{LiteRed v1.83}~\cite{Lee:2012cn,Lee:2013mka}, where the one-loop amplitude is expanded on the basis of scalar bubble ($I_2(t)$), massive ($I_3(t,m)$) and massless ($I_3(t,0)$) triangle, and two box ($I_4 (s,t)$ and $I_4 (u,t)$) integrals. The results for $\CM_4 (h,n)$ defined by the relations
\bl
\frac{\k^4 \CM_4 (h,n)}{\sbra{3} p_1 \ket{2}^{2h}} = \left. i M_4^{(1)} (p_1,k_2^{+h},k_3^{-h},p_4) \right|_{\CO(S^n)} \,,
\el
expanded in terms of master integrals
\bl
\bld
\CM_4 (h,n) &= b_1 I_4 (s,t) + b_2 I_4 (u,t) + t_1 I_3 (t,m)
\\ &\phantom{=asdf} + t_2 I_3 (t,0) + b I_2(t) \,,
\eld
\el
are presented in the ancillary file \texttt{integralcoeffs.m}~\footnote{See Supplementary Material for full integrand coefficients.} for $h = 0,1,2 $ and $n = 0,1 $, where the coefficients ($b_1$, $b_2$, $t_1$, $t_2$, and $b$) are rational functions of Mandelstam invariants and polynomials of spin Lorentz invariants [\eqc{eq:spin_Lor_inv}]. The integral coefficients were checked against overlapping known results in the literature~\cite{Bjerrum-Bohr:2016hpa,Chi:2019owc,Chen:2022yxw} and found to agree.

The eikonal phase [\eqc{eq:eik_phase_def}] is evaluated using the center of momentum (COM) frame kinematics parametrized as
\bl
\bgd
p_1^\m = (\sqrt{m^2 + \w^2}, - \vec{k} + \vec{q}/2 ) \,,\, k_2^\m = (\w, \vec{k} - \vec{q}/2) \,,
\\ p_1'^\m = (\sqrt{m^2 + \w^2}, - \vec{k} - \vec{q}/2 ) \,,\, k_2'^\m = (\w, \vec{k} + \vec{q}/2) \,,
\egd
\el
where $k := \frac{k_2 + k_2'}{2}$ and $q := k_2' - k_2$. On-shell conditions require $\vec{k} \cdot \vec{q} = 0$ and $\w^2 = \vec{k}^2 + \vec{q}^2/4$. The $\hbar$ factors are restored through the rules~\cite{Kosower:2018adc,Maybee:2019jus,Cristofoli:2021vyo,Chen:2022yxw}
\bl
G \to \frac{G}{\hbar} \,,\, S^\m \to \frac{S^\m}{\hbar} \,,\, \w \to \hbar \w \,,\, \vec{k} \to \hbar \vec{k} \,,\, \vec{q} \to \hbar \vec{q} \,,
\el
and the classical expansion parameters are chosen as
\bl
\left\{ G m |\vec{q}| \sim \frac{Gm}{b} \,,\, \frac{|\vec{q}| |\vec{S}|}{m} \sim \frac{S}{m b} \,,\, \frac{|\vec{q}|}{\w} \sim \frac{1}{\w b} = \frac{\l}{b} \right\} \,,
\el
where $b = |\vec{b}|$ is the impact parameter and $\l := \w^{-1}$ is the wavelength of the massless particle. Only the leading terms in $(\w b)^{-1}$ are considered, limiting the analyses to the geometric optics regime~\cite{Cristofoli:2021vyo,Chen:2022yxw}. $\hbar = 1$ is used unless explicit $\hbar$ counting is necessary. In the COM frame, the components of the covariant spin vector are
\bl
S^\m = \left( \frac{\vec{p}_1 \cdot \vec{S}}{m} , \vec{S} + \frac{\vec{p}_1 (\vec{p}_1 \cdot \vec{S})}{m(m + E_1)} \right) \,,
\el
where $E_1 = \sqrt{m^2 + \w^2}$ and $\vec{p}_1 = - \vec{k} + \vec{q}/2$. The following basis of spin Lorentz invariants is used in the eikonal phase where $n_\m = \e_{\a\b\g\m} p_1^\a k^\b q^\g$:
\bl
\bld
(k \cdot S) &= - \left\{ \frac{m+\w}{m} (\vec{k} \cdot \vec{S}) - \frac{\w (\vec{q} \cdot \vec{S})}{2m} \right\} [1 + \CO(\hbar^2)] \,,
\\ (q \cdot S) &= - (\vec{q} \cdot \vec{S} ) [1 + \CO(\hbar^2)] \,,
\\ (n \cdot S) &= (m + \w) \left\{ \vec{q} \cdot (\vec{k} \times \vec{S}) \right\} [1 + \CO(\hbar^2)] \,.
\eld \nn
\el
The expressions were expanded to first subleading order in $\hbar$. The little group factor $\sbra{2'} p_1 \ket{2}$ is
\bl
\bld
\sbra{2'} p_1 \ket{2} &= \sbra{2} p_1 \ket{2'} = \sqrt{\text{tr}_-(\slashed{k_2}\slashed{p_1}\slashed{k_2'}\slashed{p_1})}
\\ &= 2 (m+\w)\, \w \left[ 1 + \CO((q/\w)^2 , \hbar^2)\right] \,.
\eld
\el

The $\e^{-1}$ pole subtracted scalar integrals ($D = 4 - 2\e$) are used when converting to IPS, where $t = - \vec{q}^2$ and relevant terms for leading quantum corrections were kept. The integrals were taken from Ref.\cite{Ellis:2007qk}.
\bl
\bld
I_2 (t) &= \frac{- i}{16 \pi^2} \log (\vec{q}^2) \,,
\\ I_3 (t,0) &= \frac{-i}{32 \pi^2} \frac{\log^2 (\vec{q}^2)}{\vec{q}^2} \,,
\\ I_3 (t,m) &= \frac{-i}{32} \left( \frac{1}{m \sqrt{\vec{q}^2}} + \frac{\log(\vec{q}^2)}{\pi^2 m^2} \right) \,,
\\ I_4 (s,t) + I_4 (u,t) &= \frac{-(m-\w)}{16 \pi m^2 \w} \frac{\log(\vec{q}^2)}{\vec{q}^2} \,,
\\ I_4 (s,t) - I_4 (u,t) &= \frac{- 1}{16 \pi m \w} \frac{\log(\vec{q}^2)}{\vec{q}^2} \,.
\eld
\el
The table for IPS Fourier transform is
\bl
\bld
\int \hskip-4pt \frac{d^2 q}{(2\pi)^2} e^{i \vec{q} \cdot \vec{b}} \, \frac{1}{\vec{q}^2} &= - \frac{\log(b)}{2 \pi} \,,
\\ \int \hskip-4pt \frac{d^2 q}{(2\pi)^2} e^{i \vec{q} \cdot \vec{b}} \, \frac{1}{\sqrt{\vec{q}^2}} &= \frac{1}{2 \pi b} \,,
\\ \int \hskip-4pt \frac{d^2 q}{(2\pi)^2} e^{i \vec{q} \cdot \vec{b}} \, \log(\vec{q}^2) &= - \frac{1}{\pi b^2} \,,
\\ \int \hskip-4pt \frac{d^2 q}{(2\pi)^2} e^{i \vec{q} \cdot \vec{b}} \, \frac{\log(\vec{q}^2)}{\vec{q}^2} &= \frac{\log^2(b)}{2 \pi} \,,
\\ \int \hskip-4pt \frac{d^2 q}{(2\pi)^2} e^{i \vec{q} \cdot \vec{b}} \, \frac{\log^2(\vec{q}^2)}{\vec{q}^2} &= - \frac{2 \log^3(b)}{3 \pi} \,,
\eld
\el
where regulator-dependent terms [e.g. $\log(4 \pi)$, $\g_E$, $\z_n$, etc.] were dropped from the $\CO(\e^0)$ terms computed from the master integral eq.(3.14) of Ref.\cite{Kim:2020cvf} ($D - 2 = 2 - 2\e$).


\section{Results}
The eikonal phase [\eqc{eq:eik_phase_def}] is decomposed into the average part $\bar{\delta}$ and the GFR angle $\a$ [\eqc{eq:alpha_def}] as
\bl
\delta_{\pm \to \pm} = \bar{\delta} \mp |h| \a \,,
\el
which is further decomposed into five pieces,
\bl
\delta = \delta^{\topbot} + \delta^{\Box-\topbot^2} + \delta^{\bigtriangleup} + \delta^{\bigtriangledown} + \delta^{\bub} \,,
\el
where $\delta$ stands for $\bar{\delta}$, $\a$, or $\delta_{\pm}$. The decomposition is based on the contributing integral; $\delta^{\topbot}$ is the tree, $\delta^{\bigtriangleup}$($\delta^{\bigtriangledown}$) is the massive(massless) triangle, $\delta^{\bub}$ is the massless bubble, and $\delta^{\Box-\topbot^2}$ is the remainder that captures failure of exponentiation by the box contributions, e.g.~\footnote{Since helicity non-preserving amplitudes are present at tree level for graviton scattering, the eikonal should rather be treated as a matrix with $\delta_{\pm}$ as diagonal entries and remainder terms should be calculated using matrix algebra~\cite{Camanho:2014apa,AccettulliHuber:2020oou}. Separate treatment of $\delta_+$ and $\delta_-$ is justified as helicity non-preserving off-diagonal terms are suppressed at the considered order in $(\w b)^{-1}$~\cite{Chen:2022yxw}.}
\bl
i \delta_{+}^{\Box-\topbot^2} = \left[ \int \hskip-4pt \frac{d^2 q}{(2\pi)^2} \frac{\k^4 [b_1 I_4 (s,t) + b_2 I_4 (u,t)]}{\sbra{2} p_1 \ket{2'}^{2h}} \right] - \frac{[\delta_{+}^{\topbot}]^2}{2!} \,. \nn
\el
The remainder term is also known as the quantum remainder~\cite{KoemansCollado:2019ggb,DiVecchia:2021bdo,Haddad:2021znf}, but is known to contribute classically at first subleading order in $(\w b)^{-1}$~\cite{Chen:2022yxw}. When writing the eikonal phase, it is convenient to define the vectors
\bl
\hat{k} := \frac{\vec{k}}{\w} \,,\, \vec{a} := \frac{\vec{S}}{m} \,,\, 
\vec{d} := \frac{\vec{k} \times \vec{S}}{m \w} = \hat{k} \times \vec{a} \,.
\el
The vector $\hat{k}$ has unit length to the considered approximation order, is directed along the massless particle's propagation direction, and is orthogonal to the IPS.

The tree eikonal phase from \eqc{eq:treeComp} is independent of helicity,
\bl
\bar{\delta}^{\topbot} = \delta_{\pm}^{\topbot} = - 4 G (m + \w) \, \w \left[ \log(b) + \frac{\vec{d} \cdot \vec{b}}{b^2} \right] \,,
\el
and has quantum terms~\footnote{As a curiosity, it is remarked that the quantum terms can be absorbed by redefining $m \to m' = m + \w$, cancelling the quantum remainder terms of the spinless sector at the same time. This may have connections to the difference between the on-shell mass and the classical mass~\cite{Brandhuber:2021eyq}. However, the shift fails to cancel the quantum remainder terms of spin-linear sector.} which did not appear in the previous study, traced back to different IPS projection condition $P(q)$ and different definition of $\w$~\cite{Chen:2022yxw}.

The remainder and massive triangle contributions are found to be universal; they are the same for $h = 0,1,2$.
\bl
i \bar{\delta}^{\Box-\topbot^2} &= \frac{2 G^2 m^2}{b^2} \left[ 1 - \frac{6 ( \vec{d} \cdot \vec{b})}{b^2} + \frac{2 \w \hbar}{m} \left[ 1 - \frac{4 ( \vec{d} \cdot \vec{b})}{b^2} \right] \right] \,, \nn
\\ i \a^{\Box-\topbot^2} &= \frac{8 G^2 m^2 (\hat{k} \cdot \vec{a})}{\w b^4} \left( 1 + \frac{\w \hbar}{m} \right) \,, \nn
\\ \bar{\delta}^{\bigtriangleup} &= \frac{15 \pi G^2 m^2 \w}{4b} \left( 1 + \frac{\w \hbar}{m} \right) \left[ 1 - \frac{4}{3} \frac{(\vec{d} \cdot \vec{b})}{b^2} \right] \nn
\\ &\phantom{=a} - \frac{15}{2\pi} \left(\frac{G m}{b}\right)^2 \left( \frac{\w \hbar}{m} \right) \left[ 1 - \frac{8}{3} \frac{(\vec{d} \cdot \vec{b})}{b^2} \right] \,,
\\ \a^{\bigtriangleup} &= \frac{5 \pi G^2 m^2 (\hat{k} \cdot \vec{a})}{4 b^3} \left[ 1 - \frac{8}{\pi^2} \frac{\hbar}{m b} \right] \,.
\el
The remainder terms are purely imaginary, and the classical remainder terms are $(\w b)^{-1}$ subleading compared to the classical triangle terms~\cite{Chen:2022yxw}. Since it is unclear how to interpret the imaginary terms, the remainder terms are neglected in the final analysis.

The remaining contributions differ for external scalar($\varphi$), photon($\g$), and graviton($h$). The massless triangle contributions have a universal scaling
\bl
\bar{\delta}_{X}^{\bigtriangledown} &= \frac{8 G^2 m \w \hbar}{\pi b^2} \left[ \frac{( \vec{d} \cdot \vec{b})}{b^2} + \log(b) \left[ 1 -\frac{2 ( \vec{d} \cdot \vec{b})}{b^2} \right] \right] c_{1,X}^{\bigtriangledown} \,,
\\ \a_{X}^{\bigtriangledown} &= \frac{8 G^2 m \hbar(\hat{k} \cdot \vec{a})}{\pi b^4} \left[ - 1 + \log(b) \right] c_{2,X}^{\bigtriangledown} \,,
\el
where
\bl
c_{1,\varphi}^{\bigtriangledown} &= 3 \,,\, && & c_{1,\g}^{\bigtriangledown} &= 3 \,, && & c_{1,h}^{\bigtriangledown} &= 4 \,,
\\ & && & c_{2,\g}^{\bigtriangledown} &= 4 \,,\, && & c_{2,h}^{\bigtriangledown} &= 3 \,,
\el
and $X = \varphi, \g, h$ denotes particle species. The bubble contributions are not as orderly;
\bl
\bar{\delta}_{X}^{\bub} &= \frac{G^2 m \w \hbar}{ \pi b^2} c_{3,X}^{\bub} + \frac{G^2 m \w \hbar (\vec{d} \cdot \vec{b})}{\pi b^4} c_{4,X}^{\bub} \,,
\\ \a_{X}^{\bub} &= \frac{G^2 m \hbar(\hat{k} \cdot \vec{a})}{\pi b^4} c_{5,X}^{\bub} \,.
\el
The coefficients are found to be;
\bl
c_{3,\varphi}^{\bub} &= \frac{3}{10} \,,\, && & c_{3,\g}^{\bub} &= - \frac{161}{30} \,,\, && & c_{3,h}^{\bub} &= - \frac{29}{2} \,,
\\ c_{4,\varphi}^{\bub} &= \frac{52}{5} \,,\, && & c_{4,\g}^{\bub} &= \frac{326}{15} \,,\, && & c_{4,h}^{\bub} &= 40 \,,
\\ & && & c_{5,\g}^{\bub} &= - \frac{364}{15} \,,\, && & c_{5,h}^{\bub} &= -26 \,.
\el

Collecting the results ($\a^{\bigtriangleup}$, $\a^{\bigtriangledown}$, and $\a^{\bub}$) yields the GFR angle \eqc{eq:Faraday_rot_ang_0}, or \eqc{eq:Faraday_rot_ang_3} in dimensionless variables, after restoring reference scales ($b_0$ or $\th_0$) in the logarithms. The difference of the GFR angle between photons and gravitons is
\bl
\bld
\a_\g - \a_h &= - \left( \frac{94}{15} - 8 \log \frac{b}{b_0} \right) \frac{G^2 m \hbar (\hat{k} \cdot \vec{a})}{\pi b^4} 
\\ &= - \left( \frac{47}{240} + \frac{1}{4} \log \frac{\th}{\th_0} \right) \frac{m_{\text{Pl}}^2}{m^2} (\hat{k} \cdot \vec{\chi}) \, \th^4 \,.
\eld \label{eq:Faraday_rot_ang_4}
\el
While the remainder term $\a^{\Box-\topbot^2}$ is nonzero, the difference [\eqc{eq:Faraday_rot_ang_4}] is independent of it because its contribution is universal.

\section{Discussion}
Measuring classical frame-dragging effect is already a technological challenge~\cite{Everitt:2011hp}, and measuring its quantum corrections does not seem to be possible in the near future if not impossible [for reference, $(m_{\text{Pl}}/M_\odot)^2 \sim 10^{-78}$ for solar mass $M_\odot$]. Nevertheless, the quantum corrections to the GFR angle are interesting from a theoretical perspective, as the result provides a window into how the equivalence principle\textemdash the founding principle of GR which can be bootstrapped from physical requirements~\cite{Weinberg:1964ew}\textemdash should be understood when quantum loops are present.

In classical geometric optics approximation the GFR angle is computed by integrating the Levi-Civita connection along the ray's trajectory~\cite{Ishihara:1987dv}, which is independent of particle's species and can be understood as a manifestation of the equivalence principle~\cite{Li:2022izh,Chen:2022yxw}. But in the quantum regime different particle species experience different polarization rotation, which can be attributed to; the difference in the trajectories~\cite{Bjerrum-Bohr:2014zsa}, the difference in the frame-dragging rate, or both. Any of the suggested options are in tension with the usual formulation of the equivalence principle, which demands that massless particles travel on a universal trajectory and experience a universal frame-dragging rate.

Moreover, the difference of the GFR angle [\eqc{eq:Faraday_rot_ang_4}] is robust against redefinitions; it is invariant under the mass redefinition $m \to m \, (1 + \# \frac{\w\hbar}{m})$ or the impact parameter redefinition $b \to b \, (1 + \# \frac{Gm}{b} + \# \frac{\hbar}{mb} )$, where $\#$ are $\CO(1)$ $c$ numbers. The former is relevant since on-shell mass and classical mass can be different~\cite{Brandhuber:2021eyq}. The latter is relevant since the definition of $b$ may change due to geometry of the kinematics~\cite{Bern:2020gjj}, and since it is ill defined at the wave-packet scale ($\frac{\hbar}{m}$ is the Compton wavelength).

However, it would be premature to conclude from the findings that the equivalence principle is violated, as the tension may be resolved under scrutiny. One possibility is that the difference [\eqc{eq:Faraday_rot_ang_4}] is a tidal effect induced by the finite size of the wave-packet, for which the equivalence principle need not hold. The other possibility is that the equivalence principle only constrains outcomes of experiments, which are subject to the resolving power of the experimental apparatuses.

For the former possibility, the difference should depend on the size of the wave-packet, which in turn is determined by the massless particle's wavelength $\l = \w^{-1}$. While the quantum suppression factor $\frac{\hbar}{m b}$ in \eqc{eq:Faraday_rot_ang_4} does not depend on $\l$, the wavelength dependence could be hidden in the reference scale $b_0$. For the latter possibility, the difference should be compared with the theoretical bound on the resolution for the deflection angle and the polarization direction. Understanding how the reference scale ($b_0$ or $\th_0$) of the logarithms is determined would be necessary for exploring any of the considered possibilities, and a more thorough analysis is needed before a definite conclusion could be made. Whether the conclusions of this work, which are based on calculations of a single quantum of electromagnetic/gravitational wave scattering from a point source, still apply for classical waves described by coherent states~\cite{Cristofoli:2021jas} would be another subject for future studies.


\begin{acknowledgements}
The author would like to thank 
Manuel Accettulli Huber, Andreas Brandhuber, Gang Chen, Wei-Ming Chen, Stefano De Angelis, Yu-Tin Huang, and Gabriele Travaglini
for discussions and valuable feedback, and especially Manuel Accettulli Huber for sharing calculations of Ref.\cite{AccettulliHuber:2020oou} as an example of using the package \textsc{LiteRed}. The author would also like to thank Roman N. Lee for help on using the package \textsc{LiteRed}. The author was supported by the Science and Technology Facilities Council (STFC) Consolidated Grant ST/T000686/1 \textit{``Amplitudes, Strings and Duality”}. The author thanks KITP for their hospitality during the stay at the programme \textit{``High-Precision Gravitational Waves''}, where this work was initiated. This research was supported in part by the National Science Foundation under Grant No. NSF PHY-1748958.

\end{acknowledgements}

\bibliography{Ref.bib}

%

\end{document}